\documentclass[10pt,a4paper]{article}
\usepackage{opex3}
\usepackage{cite}
\usepackage{subfigure}

\begin{document}

\title{Towards a picosecond transform-limited nitrogen-vacancy based single photon source}

\author{Chun-Hsu Su, Andrew D. Greentree, and Lloyd C. L. Hollenberg}

\address{Quantum Communications Victoria, School of Physics, \\ The University of Melbourne, Victoria 3010, Australia}

\email{chsu@ph.unimelb.edu.au} 

\begin{abstract*}
We analyze a nitrogen-vacancy (NV$^{-}$) colour centre based single photon source based on cavity Purcell enhancement of the zero phonon line and suppression of other transitions. Optimal performance conditions of the cavity-centre system are analyzed using Master equation and quantum trajectory methods. By coupling the centre strongly to a high-finesse optical cavity [$Q\sim\mathcal{O}(10^4-10^5)$, $V\sim\lambda^3$] and using sub-picosecond optical excitation the system has striking performance, including effective lifetime of 70~ps, linewidth of 0.01~nm, near unit single photon emission probability and small [$\mathcal{O}(10^{-5})$] multi-photon probability.
\end{abstract*}

\ocis{(230.0230) Optical devices; (230.6080) Sources; (160.4760) Optical properties; (160.2220) Defect-center materials} 


\section{Introduction}
A source that produces single photons on demand is an invaluable tool for 
precision optical measurement~\cite{polzik92, giovannetti04} and is a crucial building block
for many quantum computing and communication applications. For an attenuated laser,
the number of photons per pulse follows a Poisson distribution and the multi-photon probability
becomes negligible only in the limit of small mean photon number at the expense of the single-photon probability.
In quantum computing, using single photons to store and transport quantum information is natural as information
can be easily encoded and manipulated over photonic degrees of freedom e.g. polarization.
In linear optical quantum computing (LOQC)~\cite{knill01, kok07}, it is straightforward to perform single qubit operations on photons
with elementary optical components and projective measurements generate photon-photon interactions. In quantum communication,
single photon sources can be used for unconditionally secure quantum key distribution (QKD) protocols~\cite{waks02a, waks02b}. To date, single photon generation has been demonstrated with a variety of single
quantum emitters such as atoms~\cite{kuhn02, mckeever04, barquie05, hijlkema07}, ions~\cite{keller04}, molecules~\cite{brunel99, lounis00}, diamond colour centres~\cite{kurtsiefer00, gaebel04, wang06, wu07} and semiconductor quantum dots~\cite{englund05, kako06, hennessy07, shields07}.

Diamond defects hold promise as a platform for solid-state quantum optical and
quantum computing applications~\cite{greentree06a}, and
for the study of condensed-matter analogues~\cite{hartmann06, greentree06b, angelakis07}. One of the most
well-studied systems, garnering much attention lately, is the optically active negatively-charged
nitrogen-vacancy defect (NV$^{-}$ centre). The centre consists of a substitutional nitrogen
atom and an adjacent vacancy in the carbon lattice. It forms naturally or may be
engineered within a diamond matrix using techniques such as single ion implantation~\cite{meijer05, jamieson05, rabeau06}
or chemical vapour deposition~\cite{rabeau07}. It has a combination of remarkable
properties that render it a suitable single photon source candidate. These include robustness against photobleaching,
structural stability at room temperature and demonstrated antibunching, which is the hallmark of a single photon source~\cite{kurtsiefer00}.
The centre has also been used to realize Wheeler's delayed-choice experiment~\cite{jacques07}.
However, the centre has a relatively long photoluminescence lifetime of 11.6~ns and
broad spectral width of 150~nm~\cite{manson06}, which is not optimal for daylight or optical fibre operation in QKD~\cite{beveratos02, alleaume04}.
Furthermore, the photons are not time-bandwidth limited (or indistinguishable) for the purpose of LOQC where
photon indistinguishability is crucial for Hong-Ou-Mandel interference~\cite{hom87} and hence quantum gates~\cite{rohde05}.

Preparing the centre in a high-finesse quantum cavity offers a solution to these problems.
Cavity quantum electrodynamics has been shown to induce a single-photon
Kerr nonlinearity~\cite{turchette95} and assist quantum gate operation~\cite{duan04, duan05}.
Under strong photonic confinement, the quantum emitter (the centre in this case)
in the cavity interacts coherently with photon states with the effect of modifying
photon-emission dynamics. As a result, we show that the spectral
properties of the centre can be improved to fulfill the stringent criteria for quantum information applications.
Additionally, the emission can be directed into an application or experiment as desired.
A suitable cavity is the planar photonic-band-gap (PBG) cavity that
defines an excellent cavity with small mode volume
(of order one cubic wavelength) and low loss that provides strong centre-photon coupling~\cite{akahane03, song05, noda07}
suitable for our purposes.

Advances in fabrication techniques are nearing the stage where
preparing a single crystal diamond with PBG structures using lithography
and lift-off~\cite{olivero05, tomlijenovichanic06, baldwin06, bayn07, wang07} and placing an individual NV centre in the centre of the cavity
using ion implantation techniques may be possible. The latter technique permits locating the centre to achieve full
emission enhancement. We note that cavities have been used to enhance the emission of
quantum-dots~\cite{englund05, hennessy07} and atoms~\cite{hijlkema07} for photon generation.
Our study is made in a similar spirit, but a diamond-based device has the advantages of robustness against
the environmental noise combined with simplicity of its setup. Here, without loss of generality, we theoretically study the effects of
placing the centre within a high-finesse single-moded diamond PBG cavity.
We establish the operation criteria (cavity specification
and excitation scheme) for an efficient cavity-centre based single photon source.

\section{Theory}
\begin{figure}[tb!]
\centering\includegraphics[width=0.8\textwidth]{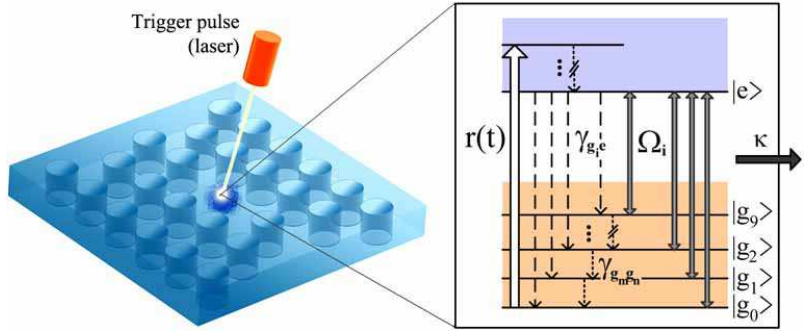}
\caption{Theoretical model of a cavity-centre system for single photon generation: the NV centre is modelled as a multi-level atom with a single excited state $|e\rangle$
and a ground state with vibrational sublevels $\{|g_j\rangle\}$.
The centre is pumped with an external classical field $r(t)$ (white arrow) acting as the trigger pulse, the transition $|e\rangle-|g_i\rangle$
is coupled to a lossy single-modal cavity with coupling strength $\Omega_i$ (grey) and cavity decay rate
$\kappa$ (black).
$\gamma_{g_ie}$ (dashed) is the atomic decay rate for radiative transitions $|e\rangle-|g_i\rangle$ while
$\gamma_{g_mg_n}$ (dotted) are that for the non-radiative phononic transitions $|g_n\rangle-|g_m\rangle$.}
\label{fig:NVModel}
\end{figure}

A model of an NV$^{-}$ centre has been proposed as a vibronic system with
ground ($g$) and excited ($e$) electronic states, each given by a series of vibrational
sublevels $|g_i\rangle$ and $|e_j\rangle$ respectively, where $i$ and $j$ label the
vibrational states~\cite{davies76}. Its emission spectra consists of several phonon lines $i$PL,
corresponding to the transition $|e_0\rangle-|g_i\rangle$. For simplicity, $|e_0\rangle$ is denoted
as $|e\rangle$ from now on. Phononic relaxation of the excited state is much faster than the radiative relaxation
to the ground state and the photonic transition probabilities were calculated by Davies and Hamer~\cite{davies76}
under the WKB approximation. 
Because of the rapid phononic transitions, we treat the centre as an atomic system with a single
excited state $|e\rangle$ and a ground state with ten sublevels $\{|g_i\rangle\}$ (Fig. \ref{fig:NVModel}).
The fluorescence from the centre corresponds to
a transition from the excited spin triplet state ($^3E$) to an electron spin triplet ground state ($^3A$)
and the dynamics are influenced by the presence of a possible metastable state ($^1A$).
We have explicitly ignored the metastable state as its role in de-excitation process remains unclear~\cite{jelezko04, manson06, tamarat06b}.
Also, in the absence of strain and magnetic field, the transitions from the $m_s=0$ spin sublevel of the ground state
are found to be spin-conserving ~\cite{manson06}, hence we assume this in our treatment.

To study the transient behaviour of the cavity-centre system, we extend the basic Jaynes-Cummings
(JC) model~\cite{jaynes63, shore93}, which treats the interaction between a single-mode
electromagnetic field or cavity of resonance frequency $\omega_C$ and a two-level atom, to consider our more complex
atomic system. In the rotating-wave approximation (RWA), the JC Hamiltonian of our cavity-centre system is expressed in terms
of the atomic projection operators $\hat{\sigma}_{\alpha\beta} = |\alpha\rangle\langle \beta|$ and
the annihilation (creation) operators $\hat{a}$ ($\hat{a}^\dagger$) for single cavity mode as, ($\hbar=1$)
\begin{equation}
 \hat{\mathcal{H}}^{\rm JC} = \sum_{i=0}^{N_A-2} \omega_{g_i}\hat{\sigma}_{g_i g_i} + \omega_{e}\hat{\sigma}_{ee} + \omega_C \hat{a}^\dagger\hat{a} + \frac{1}{2}\sum_{i=0}^{N_A-2}(\Omega_{i}\hat{a}^\dagger \hat{\sigma}_{g_i e} + {\rm h.c}),
\label{eq:hamI}
\end{equation}
where $N_A = 11$ is the total number of atomic states,
$\omega_\alpha$ is the energy of the atomic level $|\alpha\rangle$ and $\Omega_i$ is
the cavity-centre coupling constant between atomic transition $|e\rangle-|g_i\rangle$. In the dipole approximation,
the coupling is $\Omega_i = d_i[\omega_C/(2\hbar\epsilon_0V)]^{1/2}$
where $d_i$ is the dipole moment of the respective transition. 
The evolution of the cavity-centre system obeys the Liouville equation of motion for the density matrix $\rho$,
\begin{equation}
\frac{d\rho}{d t} = -i[\hat{\mathcal{H}}^{\rm JC},\rho] + \sum_{j=0}^{N_A-2}\gamma_{g_j e}\mathcal{L}[\hat{\sigma}_{g_je},\rho] +  r(t)\mathcal{L}[\hat{\sigma}_{e g_0},\rho] + \sum_{i=0}^{N_A-3}\gamma_{g_i g_{i+1}}\mathcal{L}[\hat{\sigma}_{g_i g_{i+1}},\rho] + \kappa\mathcal{L}[\hat{b}^{\dagger} \hat{a},\rho],
\label{eq:master}
\end{equation}
with the Lindbladian terms for some operator $\hat{O}$,
\begin{equation}
  \mathcal{L}[\hat{O},\rho] = \hat{O}\rho \hat{O}^\dagger - \frac{1}{2}\left(\hat{O}^\dagger \hat{O} \rho + \rho \hat{O}^\dagger \hat{O}\right).
\label{eq:lin}
\end{equation}
The spontaneous transition ($|e\rangle-|g_j\rangle$) couples to any non-cavity field modes at
the characteristic rates $\gamma_{g_je}$ and non-radiative phononic
decays from $|g_{i+1}\rangle$ to $|g_i\rangle$ with rates $\gamma_{g_i g_{i+1}}$. Here
$\hat{b}^\dagger$ represents the creation operator for electromagnetic (waveguide) mode
outside the cavity, which the cavity couples to via decay rate $\kappa$. The rate $\kappa = \omega_C/(2Q)$ is parameterized
by the quality factor of the cavity $Q$. Incoherent excitation, acting as the trigger for photon emission,
is represented by the phenomenological term with a pump absorption rate $r(t)$.
In this model, we have explicitly ignored thermal broadening by assuming a zero temperature operating
environment. Broadening can be introduced phenomenologically
for a more realistic estimate of the linewidth of the emitted wave packet.

Efficient single photon generation requires minimizing 
loss and fast outcoupling of the
excitation via the cavity channel. The appropriate regime to
optimise the output is the strong Purcell regime~\cite{law97}, 
$\kappa > \Omega_i \gg \gamma_{g_je}$, 
where the rate of coherent coupling between the centre and cavity
mode, $\Omega_i^2/\kappa$, dominates the rates $\gamma_{g_je}$ ($\forall j$) of the
incoherent coupling to the non-cavity modes. The cavity mode
is chosen to match the transition $|e\rangle-|g_i\rangle$. The
cavity loss rate $\kappa$ sets the time scale for photon outcoupling
and when greater than $\Omega_i$, suppresses the vacuum Rabi oscillations,
which otherwise lead to unwanted spectral features
on the output photon.  For a two-level atom in a cavity, altogether embedded in
a medium of refractive index $n$, the enhancement of emission into the cavity is parameterized by the
Purcell factor, the ratio of the emission
rate to the cavity mode to the unmodified rate into the non-cavity
modes~\cite{purcell46, haroche85, gerard99},
\begin{equation}
    F_{p} = \frac{3(\lambda_C/n)^3}{4\pi^2}\frac{Q}{V},
\label{eq:purcell}
\end{equation}
where $V$ is the cavity mode volume and $\lambda_C$ is the cavity wavelength.

To determine the fraction of pulse cycles that lead to useful
output, we must consider the fraction of emitted photons from the
atom that enter the desired cavity mode.  This fraction is set by
the spontaneous coupling factor $\beta = F_p/(1+F_p)$. We demand
$\beta$ to be near unity, which implies a high-$Q/V$ cavity to maximize
efficiency, but a small $Q$ for fast cavity loss. These
considerations lead to an upper bound for $Q$, or equivalently, a lower bound on
$\kappa$ given by $\kappa \geq 2\Omega_i$. Beyond this, we
enter the strong cavity regime \cite{haroche85}. In this regime, the
vacuum Rabi oscillations are not sufficiently suppressed by cavity
decay leading to an effective timing jitter or equivalently
temporal/spectral features which will degrade overall device
performance. Alternatively, if $\kappa~<~\gamma_{g_je}$ for large $Q$, the excitation will be outcoupled
as atomic decoherence, which leads to loss of photons through
non-guided modes, and manifests as an increased zero photon
probability. A rigorous treatment of
cavity-assisted emission was made in Ref.~\cite{khanbekyan07}.

\section{Analysis and results}
To investigate dynamical processes of photon generation from the cavity-centre system
described by Eq.~\ref{eq:master}, we use direct numerical integration to
study the specifications on the trigger pulse that ensures high-fidelity single photon emission and the effect of the cavity
on its characteristics. Additionally, we use the quantum trajectory approach to simulate photodetection
experiments~\cite{tian92, carmichael93}. We show that it yields results that agree with the former approach and demonstrate
bit-stream photon generation and, most importantly, antibunching with the Hanbury-Brown-Twiss (HBT) experiment.

\begin{figure}[tb!]
\subfigure{
{\includegraphics[width=0.52\textwidth]{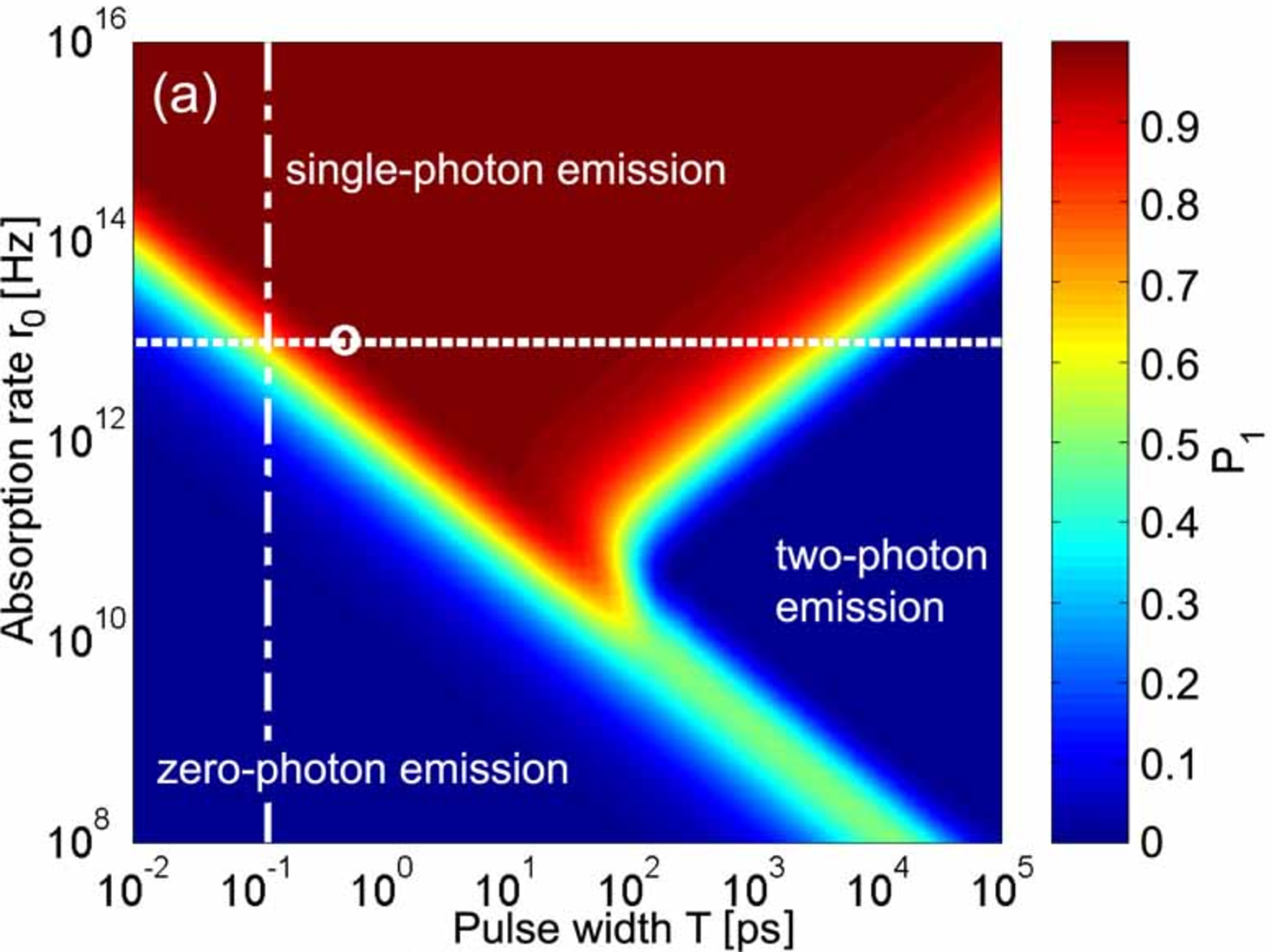}}}
\subfigure{
{\includegraphics[width=0.48\textwidth]{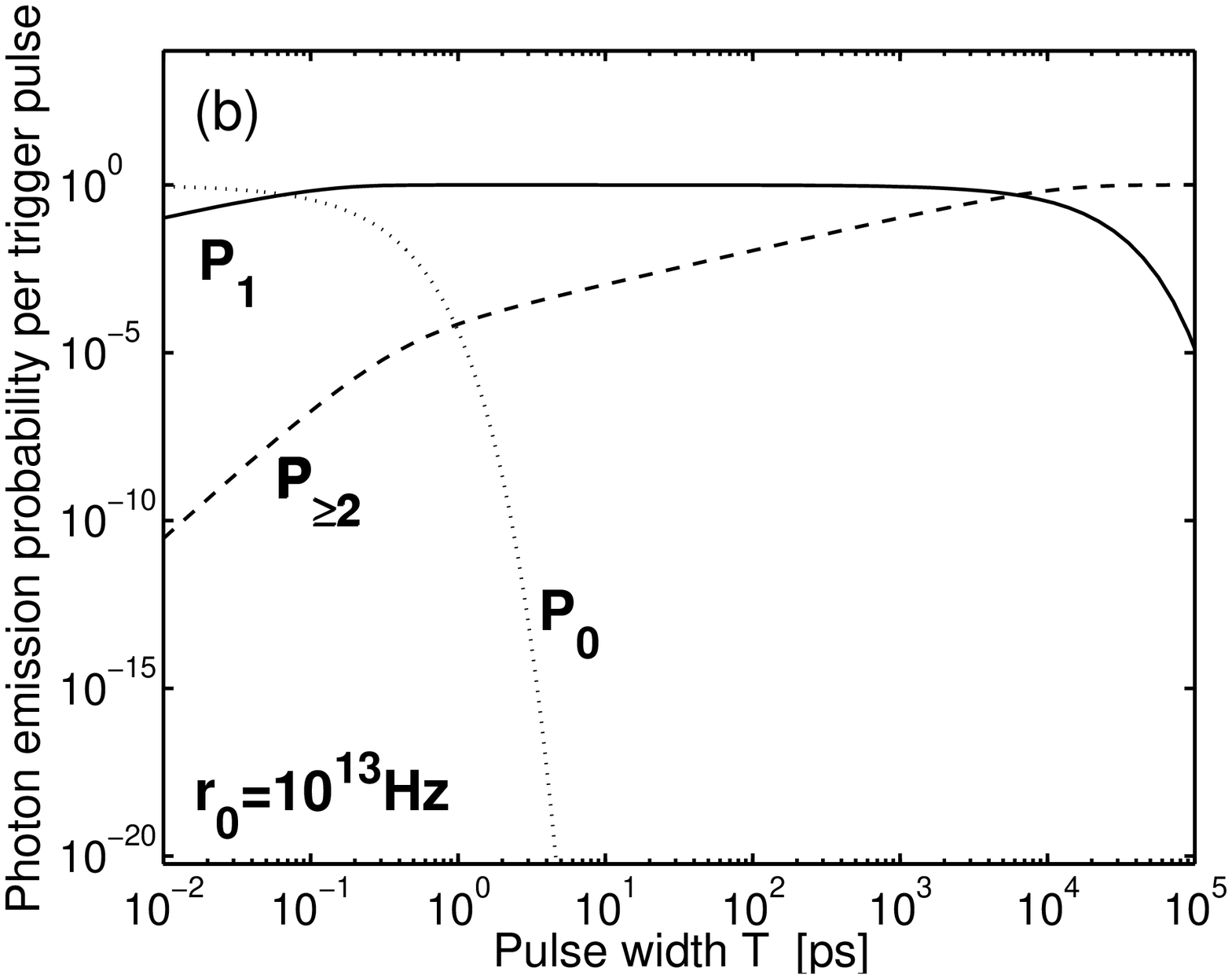}}}
\caption{
{\bf a. } Probability of the cavity-centre system ($\omega_C=\omega_{ZPL}$, $V = \lambda_{ZPL}^3$)
to emit a single photon per top-hat excitation pulse
as a function of pulse width $T$ and absorption rate $r_0$. Dash-dotted line denotes the pulse width
parameter used in Ref.~\cite{dumeige04} where the illumination
irreversibly transforms the centre into a different centre and is therefore a practical cutoff. 
Circle labels the parameters [yield $P_{1} = 0.996$
and $P_{\geq2}\sim\mathcal{O}(10^{-5})$] used for the demonstration of single-photon generation with the cavity-centre system.
{\bf b. } Zero (dotted), single (solid) and multi- (dashed) photon probability as a function of
pulse width with constant absorption rate $r_0 = 10^{13}$~Hz.}
\label{fig:PrelimCal}
\end{figure}

\subsection{Determination of the pulsed excitation parameters}
The specification of the pulsed excitation scheme for the
cavity-centre system is crucial to ensure efficient and
high-fidelity single photon emission. The complete treatment of
Eq. \ref{eq:master} for this study is computationally demanding as it 
requires adopting a considerably large state space
given by $\{|\alpha\rangle\}_{\rm atomic}\otimes\{|0_C\rangle, |1_C\rangle, ..., |N_C\rangle\}_{\rm cavity}\otimes\{|0_W\rangle, |1_W\rangle, ..., |N_W\rangle\}_{\rm waveguide}$, where the state evolution is
taken to span over $N_W$ (or equivalently $N_C$) excitation states. 
`$C$' denotes the field mode in the cavity and `$W$' labels the external waveguide mode.
However, for the purposes of understanding just the
photon emission, we may reduce the system state space by setting
$N_C = 2$, $N_W = 0$ and enforcing a choice of parameters for the
pulsed excitation that limits system population to the $\leq
2$-quantum states throughout the cycle.

For the excitation, we assume a top-hat form of the trigger pulse
$r(t) = r_0$ for $t\in[0,T]$, where $r_0$ is the absorption rate and
$T$ the pulse width which must be much shorter than the photoluminescence lifetime. 
In the strong Purcell regime the centre may be
approximated as a two-level system in resonance with the cavity, and for very short
times we ignore the effects of spontaneous emission and cavity outcoupling. 
Under these approximations the
lower bound for zero photon emission probability ($P_{0}$) and the upper bound
for one ($P_{1}$) are
\begin{eqnarray}
 P_{0} & = & {\rm e}^{-r_0T}, \label{eq:probs1} \\
 P_{1} & = & 2 {\rm e}^{-r_0T}\left\{\frac{{\rm e}^{r_0T/2}[-16\Omega_{i}^2 + r_0^2 \cosh(\eta T/2)]}{\eta^2} - 1\right\}, \label{eq:probs2}
\end{eqnarray}
and upper bound for multi-photon probabilities is $P_{\geq2} = 1 - (P_{0} + P_{1})$, where $\eta = (r_0^2-16\Omega_i^2)^{1/2}$. In the limit $r_0 \gg
\Omega_i$, Eq. \ref{eq:probs2} reduces to $P_{1} = 1-\exp(-r_0T)$, but the
second order term reveals that the single photon probability
decreases with increasing pulse width according to
$\exp(-4\Omega_i^2T/r_0) - P_{0}'$, where $P_{0}' =
[2-\exp(-4\Omega_i^2T/r_0)]\exp(-r_0T)$. There is therefore a clear
trade-off between the requirement to produce a photon on demand and
that of having no more than one photon per pulse.
Fig.~\ref{fig:PrelimCal} illustrates the single photon emission probability for an
NV centre in a cavity resonant with the zero-phonon line (ZPL)
transition ($|e\rangle-|g_0\rangle$), as a function of excitation parameters. In contrast to an attenuated
laser where small $P_{\geq2}$ is
achieved in expense of $P_{1}$, the cavity-centre system
offers $P_{1} \sim \mathcal{O}(1)$ while keeping
$P_{\geq2} \leq \mathcal{O}(10^{-2})$. The mean photon number
per pulse is $\bar{n} = \sum_i i P_{i} \approx
P_{1} + 2P_{\geq2}$ if $P_{i}$ is negligible for
$i\geq3$, for these parameter choices.

\begin{figure}[tb!]
\subfigure{
{\includegraphics[width=0.5\textwidth]{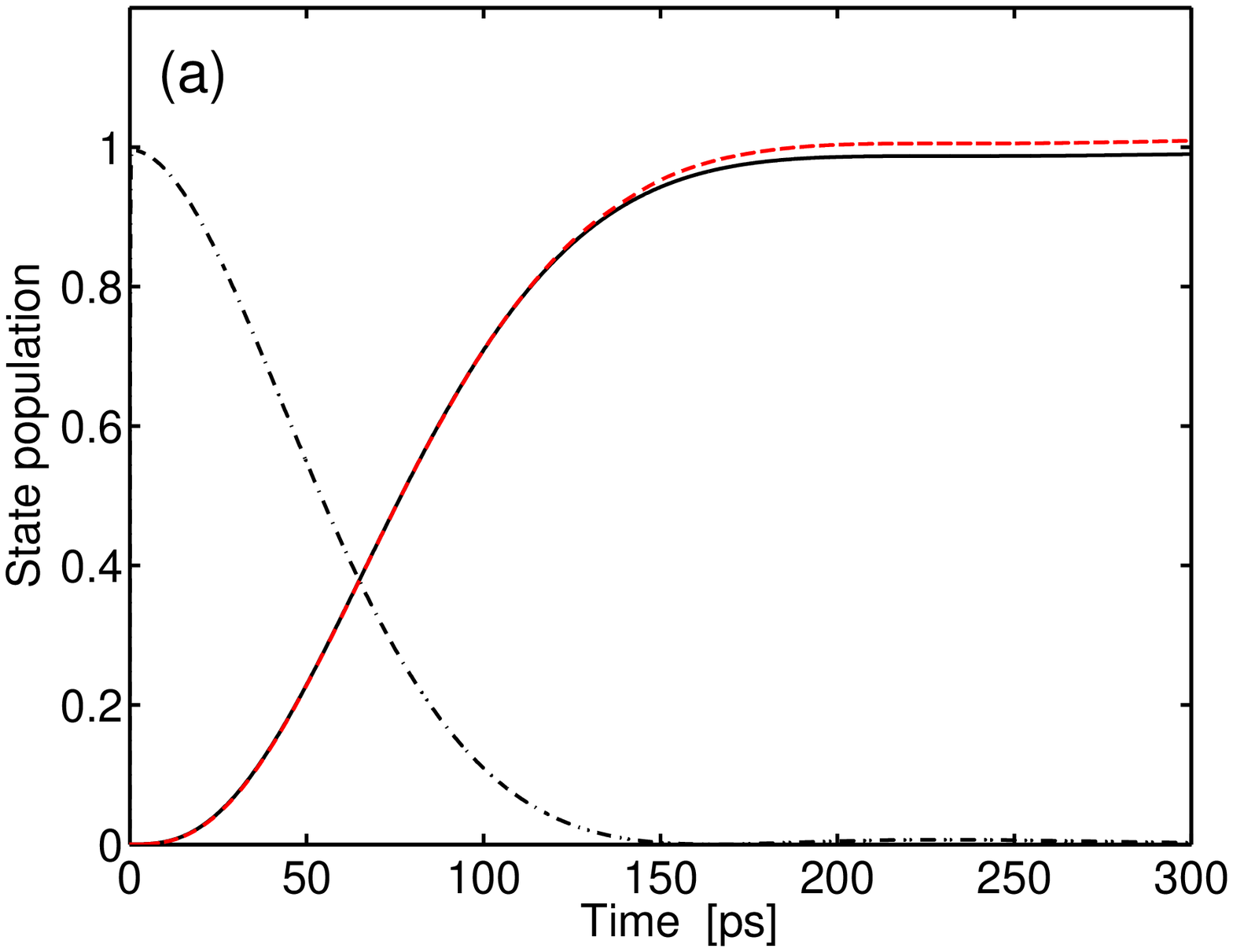}}}
\subfigure{
{\includegraphics[width=0.5\textwidth]{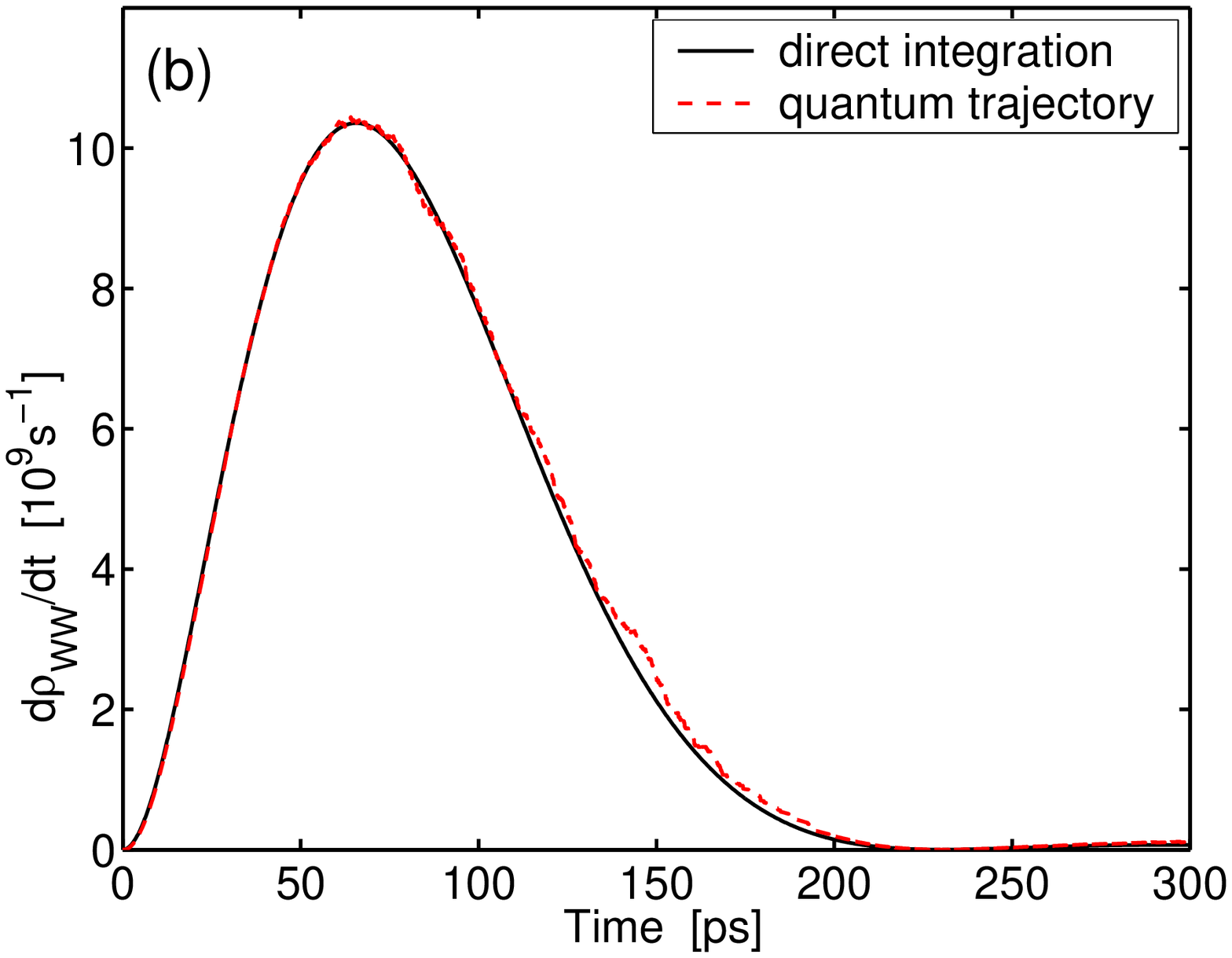}}}
\caption{Evolution of a cavity-centre system ($\omega_C=\omega_{ZPL}$, $V=\lambda_{ZPL}^3$,
$\kappa=2.5\Omega_0$) in response to a top-hat excitation pulse ($r_0 = 10^{13}$~Hz, $T = 0.56$~ps). 
{\bf a.} Population in $|e,0_C,0_W\rangle$ (the excited centre, black dash-dotted) and
$|g_0,0_C,1_W\rangle$ or $\rho_{WW}$ (the outcoupled waveguide mode, black/red solid) as a function of time. {\bf b.} Time derivatives of $\rho_{WW}$,
proportional to the output intensity, with an integrated area of 0.99 (solid) and of 1.01 (dashed red). Simulation is performed
by direct integration of Eq. \ref{eq:master} (black solid/dash-dotted) and by quantum trajectory approach as
a direct photodetection experiment (red dashed).}
\label{fig:CNVsim}
\end{figure}

The photon source must operate in the limit of large $r_0$ and maintain short $T$ to
ensure near unit single photon probability and efficient operation. However, Dumeige \textit{et al.}
showed that under intense femtosecond illumination, the centre becomes photo-ionized, resulting in blinking~\cite{dumeige04},
thus this sets the lower bound for the pulse width.

The optimal excitation parameters for the operation of cavity-centre system depend on the desired application for the
photon source. The source can be tailored by varying pulse parameters
$T$ and $r_0$ to optimize properties $P_{1}$ and $\bar{n}$ for a specific application.
We adopted technically feasible values $T = 0.56$~ps and $r_0 = 10^{13}$~Hz that yield $P_{1} = 0.996$
and $P_{\geq2} = \mathcal{O}(10^{-5})$.

\subsection{Simulation of single photon generation with the cavity-centre system}
The analysis in Sec. 3.1 allows a determination of the parameter range for single photon emission,
and this is obviously the regime in which we want our device to work. In this section, therefore,
we work in the single photon emission regime and \emph{assume} single photon
output. This affords a considerable saving in computational complexity and we thus
truncate the photonic state space to one excitation by setting $N_C = N_W = 1$. We may then solve Eq.~\ref{eq:master}
numerically to simulate the response of the cavity-centre system to a pulsed excitation,
with the result given in Fig. \ref{fig:CNVsim}.
The cavity of volume $V=\lambda_{ZPL}^3$ is chosen to maximize coupling and is in resonance with the ZPL. We
choose $Q$ of $36500$ so that $\kappa=2.5\Omega_0$.
A photon is being issued at a mean time 70~ps from the pump with excitation outcoupling into the external waveguide mode
illustrated by increasing population in $|g_0,0_C,1_W\rangle$ (or $\rho_{WW} \equiv \langle g_0,0_C,1_W|\rho|g_0,0_C,1_W\rangle$).
The integral of the derivative $\dot{\rho}_{WW}$ is near unity at 0.99, as required of
a single photon pulse. Fourier transform of the temporal profile yields an emission spectrum
centered at $\lambda_{ZPL}$ with effective linewidth of 0.01~nm.

Due to the difference between cavity-centre coupling
$\Omega_0$ and cavity outcoupling $\kappa$, the resultant photon pulse does not
take the form of a Gaussian function which is optimal for LOQC~\cite{rohde05}.
In principle, Stark tuning~\cite{tamarat06a} can be used to optimize the atom-cavity coupling to reshape
the temporal pulse profile~\cite{greentree06c, fernee07} and suppress timing jitter. Note that
there is a gentle hump at 300~ps, representing the Rabi remnant, that can be eliminated with greater cavity damping.

To simulate photodetection
experiments via the quantum trajectory approach, we again adopt a two-level model of the centre, but
truncate $N_C=4$ to allow for the possibility of multi-photon occupation and subsequent emission.
There is a good agreement between this result and that from direct integration, as shown in
Fig. \ref{fig:CNVsim}.

\begin{figure}[tb!]
\centering\includegraphics[width=8.5cm]{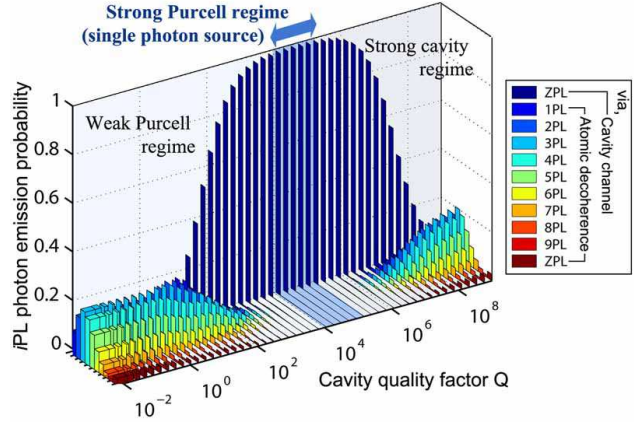}
\caption{Comparison of the probability of the cavity-centre system
($\omega_C=\omega_{ZPL}$, $V=\lambda_{ZPL}^3$)
to emit a photon of $\omega_{ZPL}$ via the cavity channel and to emit $i$PL photon via atom decoherence
as a function of cavity quality factor $Q$.}
\label{fig:CNViPL}
\end{figure}

It is instructive to observe how the probability of the system to emit a photon via the cavity and via atomic decoherence
varies under the influence of the cavity. We recalculate the emission (Fig.~\ref{fig:CNViPL}), but this time treat ($N_A$--1) vibrational ground states, and only consider one photonic excitation.
In the weak Purcell regime, the excitation is outcoupled via the atomic decoherence channel.
The relative transition probabilities of the respective phonon lines $j$PL approach the unmodified
atomic branching ratios $\gamma_{g_je}$ in the limit of small $Q$.
However, in the strong Purcell regime with $Q\sim\mathcal{O}(10^{4}-10^{5})$,
the ZPL transition is enhanced by the cavity while other transitions are suppressed accordingly.
The excitation is predominantly outcoupled via the cavity relaxation channel, representing
the optimal regime for the single-photon source. We note that such $Q$'s are technically
achievable and have been demonstrated in Ref.~\cite{akahane03, song05} in silicon, and are feasible in diamond \cite{tomlijenovichanic06}.
Finally, in the strong cavity regime, the Purcell enhancement diminishes as the time scale for cavity
relaxation becomes much longer with $\kappa<\gamma_{g_je}$.

We have also considered the possibility of enhancing the higher order phonon side bands.
Such an enhancement is attractive from the point of view of shifting the emission further into the
IR, and taking advantage of the increased dipole moments.
The presence of the vibrational sublevels and phononic decays introduces an additional
complication to the dynamics of the basic JC model. By explicitly considering a reduced
centre model with $|e\rangle$ and $|g_i\rangle$ ($i=0,1,2$ only) that couples the cavity
that is in resonance with the 1PL, we have obtained an analytical expression to the
modified decay rate by solving the master equation for its eigenfrequencies
and assuming $\Omega_0 \ll (\kappa + \gamma_{g_mg_n})$. Single excitation relaxes at an
overall damping rate,
\begin{equation}
    \gamma_{\rm overall} = \gamma_{g_0g_1} + \frac{2\Omega_0^2}{\omega_C/(2Q) + \gamma_{g_mg_n} - 2\gamma_{g_0g_1}} + \mathcal{O}(\Omega_0^4).
\end{equation}
Eq. \ref{eq:purcell} is recovered by setting $\gamma_{g_0g_1}=
\gamma_{g_mg_n} = 0$. The phononic decay reduces the coherence between the states
$|g_1,1_C\rangle$ and $|e,0_C\rangle$, inhibiting the Purcell enhancement. Hence,
to enhance the higher phonon transitions, one requires $\Omega_0$
of order $\mathcal{O}(\gamma_{g_mg_n})$ or a cavity whose modal volume is much
smaller than the scale of wavelength$^3$. Although technically demanding, we note that
sub-wavelength confinement may be possible with plasmonic cavities~\cite{chang06}.

\begin{figure}[tb!]
\subfigure{
{\includegraphics[width=0.49\textwidth]{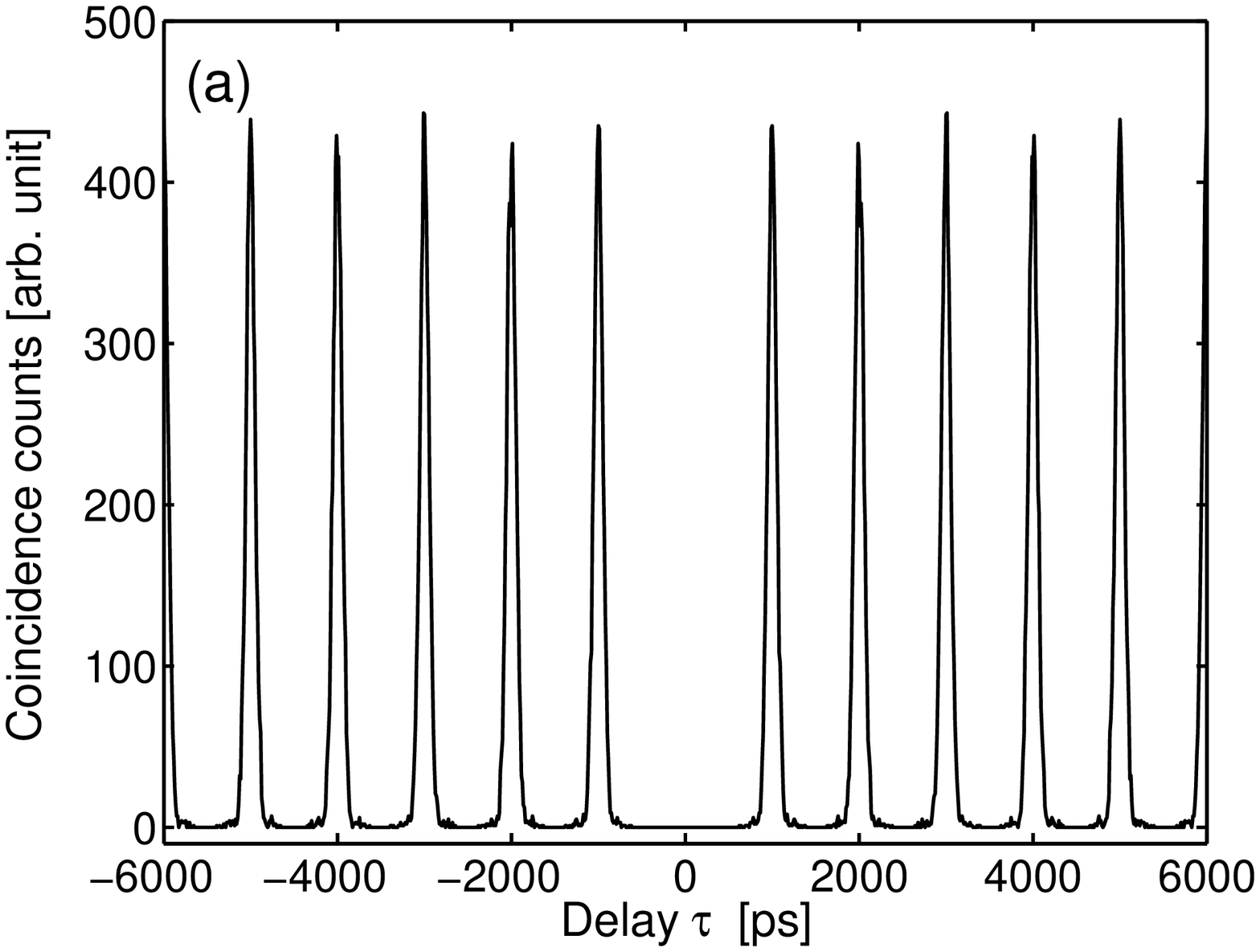}}}
\subfigure{
{\includegraphics[width=0.51\textwidth]{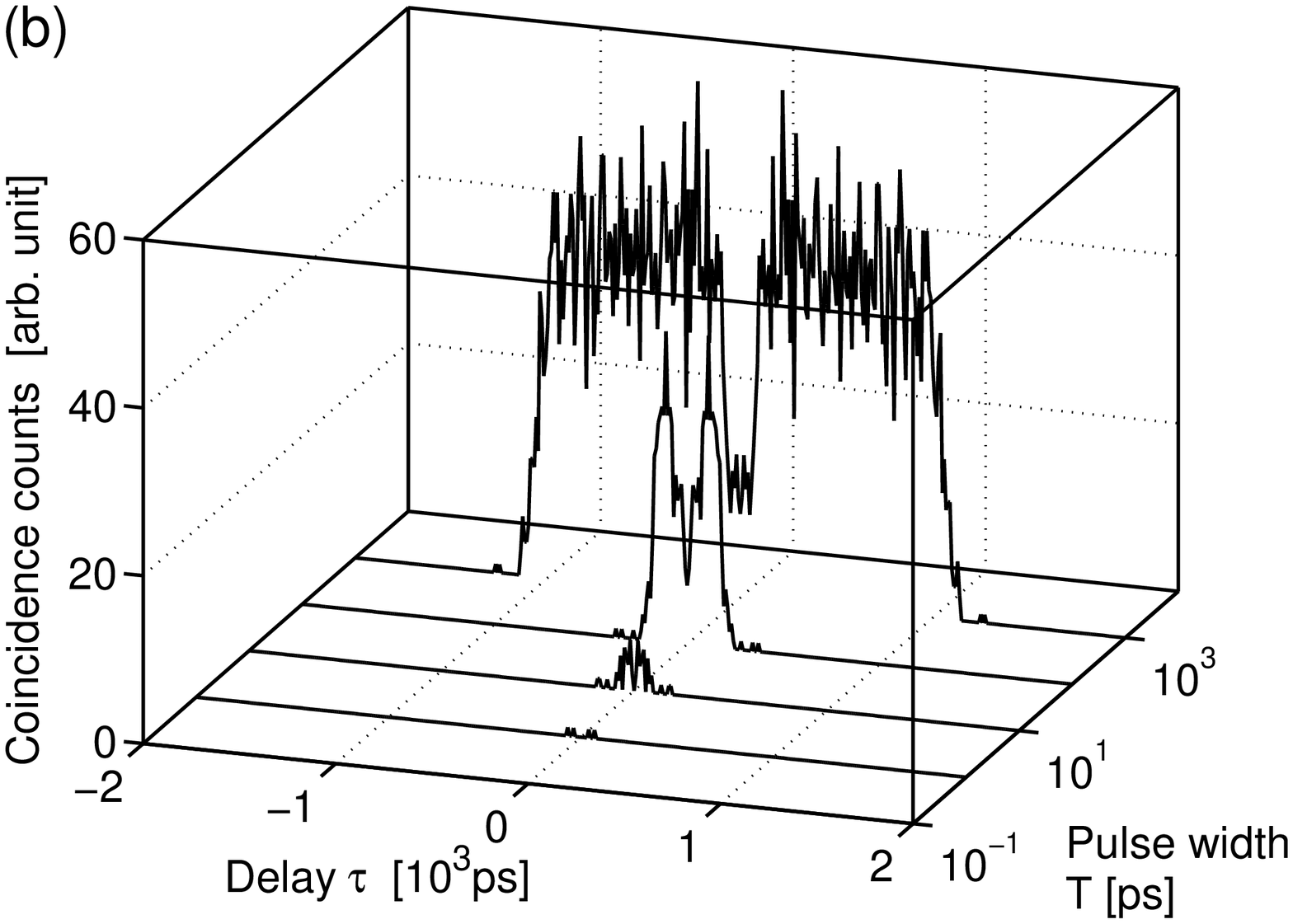}}}
\caption{Photon correlation histogram of emission from the cavity-centre
system under pulsed excitation of top-hat functional form
obtained using a HBT setup with quantum trajectory approach. The simulations
involve {\bf a.} excitation pulse of temporal width $T = 0.56$~ps
and constant absorption rate $r_0 = 10^{13}$~Hz at a repetition rate of 1~GHz for a trajectory
time of 5~$\mu$s, and {\bf b.} excitation pulse of varying temporal widths
and constant $r_0 = 10^{13}$~Hz at repetition rate of 0.5~GHz for total time 1.5~ms.}
\label{fig:Qexp}
\end{figure}

\subsection{Antibunching with a HBT setup}
Photon correlation obtained from a HBT experiment is a test for
single-photon emission. The HBT setup is simulated with a pulsed excitation
of rate 1~GHz for a simulation time of 5~$\mu$s and the correlation
histogram of emission from the cavity-centre system is shown in Fig. \ref{fig:Qexp}(a). The
experiment assumes the use of a detector with a time bin size of 10~ps.
The result shows a series of peaks separated by the clock period of the
pulse. The suppression of the coincidences observed at zero delay signifies antibunching.
More importantly, the suppression is observed during the period of a single excitation cycle for the short excitation pulse
width $T = 0.56$~ps. The single photon probability per excitation trigger,
estimated from the ratio of the number of
single photon events to total number of pulses simulated is 0.99, while multi-photon
probability is zero. With an effective lifetime of 70~ps, the system is capable of operating
at an excitation rate of 10~GHz. However, to ensure all photon pulses are well-localized within their respective time bins,
a bit-stream rate of 1~GHz is preferable.
Finally, in Fig.~\ref{fig:Qexp}(b), we show an increase in multi-photon probability with
increasing excitation pulse width. In agreement with the result from Fig.~\ref{fig:PrelimCal}(b), a considerably long simulation is needed to observe
an appreciable multi-photon probability of $\mathcal{O}(1)$\% for $T\sim10^3$~ps,
the simulation was performed over a trajectory period of 1.5~ms and resolution set to 10~ps.

\section{Conclusion}
We have studied the effect of a cavity on an NV$^-$ defect centre in enhancing its spectral
properties for the purpose of single photon generation for quantum computing and communication.
Assuming an atomic-vibronic NV model in single-mode cavity, we have shown that by coupling
the centre strongly to a high-$Q/V$ [$Q\sim\mathcal{O}(10^4-10^5)$, $V\sim$ $\lambda^{3}$]
cavity that is resonant with the ZPL and with excitation using a sub-picosecond pump, the cavity-centre system is capable of issuing a photon of wavelength 638~nm with high spectral purity.
We predict that the cavity-enhanced NV centre can have an effective lifetime of 70~ps and linewidth of 0.01~nm, in contrast with
an unmodified centre's photoluminescence lifetime of 11.6ns and spectral width of 150~nm.
Photons are emitted with near unit single photon probability of 0.99
while maintaining small multi-photon probability $\mathcal{O}(10^{-5})$, thus making it a relatively efficient triggered
photon source compared to a bare NV centre or an attenuated laser. Finally, the device can potentially operate
at a repetition rate of 1~GHz, considerably greater than demonstrated NV systems for QKD applications~\cite{beveratos02, alleaume04}.

\section*{Acknowledgments}
We thank J. H. Cole and A. M. Stephens for helpful discussions.
This project was supported by Quantum Communications Victoria (QCV), which
is funded by the Victorian Government's Science, 
Technology and Innovation (STI) initiative, and by the Australian Research Council (ARC), 
the Department of Education Science and Technology under the International Science Linkages scheme and the EU 6th Framework under the EQUIND collaboration.
ADG is the recipient of an Australian Research Council Queen
Elizabeth II Fellowship (project number DP0880466), LCLH is
the recipient of an Australian Research Council Australian Professorial
Fellowship (project number DP0770715).


\begin{thebibliography}{99}

\bibitem{polzik92} E. S. Polzik, J. Carri, and H. J. Kimble, ``Spectroscopy with squeezed light,'' \prl {\bf 68, } 3020--3023 (1992).

\bibitem{giovannetti04} V. Giovannetti, S. Lloyd, and L. Maccone, ``Quantum-enhanced measurements: Beating the standard quantum limit,'' Science\ {\bf 306, } 1330--1336 (2004).

\bibitem{knill01} E. Knill, R. Laflamme, and G. J. Milburn, ``A scheme for efficient quantum computation with linear optics,'' \nat {\bf 409, } 46--52 (2001).

\bibitem{kok07} P. Kok, W. J. Munro, K. Nemoto, T. C. Ralph, J. P. Dowling, and G. J. Milburn, ``Linear optical quantum computing,'' \rmp {\bf 79, } 135--174 (2007).

\bibitem{waks02a} E. Waks, K. Inoue, C. Santori, D. Fattal, J. Vuckovic, G. S. Solomon, and Y. Yamamoto, ``Quantum cryptography with a photon turnstile,'' \nat {\bf 420, } 762 (2002).

\bibitem{waks02b} E. Waks, C. Santori, and Y. Yamamoto, ``Security aspects of quantum key distribution with sub-Poisson light,'' \pra {\bf 66, } 042315 (2002).

\bibitem{kuhn02} A. Kuhn, M. Hennrich, and G. Rempe, ``Deterministic single-photon source for distributed quantum networking,'' \prl {\bf 89, } 067901 (2002).

\bibitem{mckeever04} J. McKeever, A. Boca, A. D. Boozer, R. Miller, J. R. Buck, A. Buzmich, and H. J. Kimble, ``Deterministic generation of single photons from one atom trapped in a cavity, '' Science\ {\bf 303, } 1992--1994 (2004).

\bibitem{barquie05} B. Barqui\'e, M. P. A. Jones, J. Dingjan, J. Beugnon, S. Bergamini, Y. Sortais, G. Messin, A. Browaeys, and P. Grangier, ``Controlled single-photon emission from a single trapped two-level atom,'' Science\ {\bf 309, } 454--456 (2005).

\bibitem{hijlkema07} M. Hijlkema, B. Weber, H. P. Specht, S. C. Webster, A. Kuhn, and G. Rempe, ``A single-photon server with just one atom,'' Nature\ Physics\ {\bf 3, } 253--255 (2007).

\bibitem{keller04} M. Keller, B. Lange, K. Hayasaka, W. Lange, and H. Walther, ``Continuous generation of single photons with controlled waveform in an ion-trap cavity system,'' \nat {\bf 431, } 1075--1078 (2004).

\bibitem{brunel99} C. Brunel, B. Lounis, Ph. Tamarat, and M. Orrit, ``Triggered source of single photons based on controlled single molecule fluorescence,'' \prl {\bf 83, } 2722--2725 (1999).

\bibitem{lounis00} B. Lounis and W. E. Moerner, ``Single photons on demand from a single molecule at room temperature,'' \nat {\bf 407, } 491--493 (2000).

\bibitem{kurtsiefer00} C. Kurtsiefer, S. Mayer, P. Zarda, and H. Weinfurter, ``Stable solid-state source of single-photons,'' \prl {\bf 85, } 290--293 (2000).

\bibitem{gaebel04} T. Gaebel, I. Popa, A. Gruber, M. Domhan, F. Jelezko, and J. Wrachtrup, ``Stable single-photon source in the near infrared,'' New\ J.\ Phys.\ {\bf 6, } 98 (2004).

\bibitem{wang06} C. Wang, C. Kurtsiefer, H. Weinfurter, and B. Burchard, ``Single photon emission from SiV centres in diamond produced by ion implantation,'' J.\ Phys.\ B:\ At.\ Mol.\ Opt.\ Phys.\ {\bf 39,} 37--41 (2006).

\bibitem{wu07} E. Wu, J. R. Rabeau, G. Roger, F. Treussart, H. Zeng, P. Grangier, S. Prawer, and J.-F. Roch, ``Room temperature triggered single-photon source in the near infrared,'' New\ J.\ Phys.\ {\bf 9,} 434 (2007).

\bibitem{englund05} D. Englund, D. Fattal, E. Waks, G. Solomon, B. Zhang, T. Nakaoka, Y. Arakawa, Y. Yamamoto, and J. Vu\v{c}kovi\'c, ``Controlling the spontaneous emission rate of single quantum dots in a two-dimensional photonic crystal,'' \prl {\bf 95, } 013904 (2005).

\bibitem{kako06} S. Kako, C. Santori, K. Hoshino, S. G\"otzinger, Y. Yamamoto, and Y. Arakawa, ``A gallium nitride single-photon source operating at 200K,'' Nature\ Materials\ {\bf 5, } 887--892 (2006).

\bibitem{hennessy07} K. Hennessy, A. Badolato, M. Winger, D. Gerace, M. Atat\"ure, S. Gulde, S. F\"alt, E. L. Hu, and A. Imamo\v{g}lu, ``Quantum nature of a strongly coupled single quantum dot-cavity system,'' \nat {\bf 445, }  896--899 (2007).

\bibitem{shields07} A. J. Shields, ``Semiconductor quantum light sources,'' Nature\ Photonics\ {\bf 1, } 215--223 (2007).

\bibitem{greentree06a} A. D. Greentree, P. Olivero, M. Draginski, E. Trajkov, J. R. Rabeau, P. Reichart, B. C. Gibson, S. Rubanov, S. T. Huntington, D. N. Jamieson, and S. Prawer, ``Critical components for diamond-based quantum coherent devices,'' J.\ Phys.:\ Cond.\ Matt. {\bf 18, } S825--S842 (2006).

\bibitem{hartmann06} M. J. Hartmann, F. G. S. L. Brand\~{a}o, and M. B. Plenio, ``Strongly interacting polaritons in coupled arrays of cavities,'' Nature\ Physics\ {\bf 2, } 849--855 (2006).

\bibitem{greentree06b} A. D. Greentree, C. Tahan, J. H. Cole, and L. C. L. Hollenberg, ``Quantum phase transitions of light,'' Nature\ Physics {\bf 2, } 856--861 (2006).

\bibitem{angelakis07} D. G. Angelakis, M. F. Santos, and S. Bose, ``Photon-blockade-induced Mott transitions and XY spin models in coupled cavity arrays,'' \pra {\bf 76, } 031805(R) (2007).

\bibitem{meijer05} J. Meijer, B. Burchard, M. Domhan, C. Wittmann, T. Gaebel, I. Popa, F. Jelezko, and J. Wrachtrup, ``Generation of single color centers by focused nitrogen implantation,'' \apl {\bf 87, } 261909 (2005).

\bibitem{jamieson05} D. N. Jamieson, C. Yang, T. Hopf, S. M. Hearne, C. I. Pakes, S. Prawer, M. Mitic, E. Gauja, S. E. Andreson, F. E. Hudson, A. S. Dzurak, and R. G. Clark, ``Controlled shallow single-ion implantation in silicon using an active substrate for sub-20-keV ions,'' \apl {\bf 86, } 202101 (2005).

\bibitem{rabeau06} J. R. Rabeau, P. Reichart, G. Tamanyan, D. N. Jamieson, S. Prawer, F. Jelezko, T. Gaebel, I. Popa, M. Domhan, and J. Wrachtrup, ``Implantation of labelled single nitrogen vacancy centres in diamond using $^{15}$N,'' \apl {\bf 88, } 23113 (2006).

\bibitem{rabeau07} J. R. Rabeau, A. Stacey, A. Rabeau, F. Jelezko, I. Mirza, J. Wrachtrup, and S. Prawer, ``Single nitrogen vacancy centers in chemical vapor deposited diamond nanocrystals,'' Nano\ Letters\ {\bf 7, } 3433--3437 (2007).

\bibitem{jacques07} V. Jacques, E. Wu, F. Grosshans, F. Treussart, P. Grangier, A. Aspect, and J.-F. Roch, ``Experimental realization of Wheeler's delayed-choice gedanken experiment,'' Science\ {\bf 315, } 966--968 (2007).

\bibitem{manson06} N. B. Manson, J. P. Harrison, and M. J. Sellars, ``Nitrogen-vacancy center in diamond: Model of the electronic structure and associated dynamics,'' \prb {\bf 74, } 104303 (2006).

\bibitem{beveratos02} A. Beveratos, R. Brouri, T. Gacoin, A. Villing, J.-P. Poizat, and P. Grangier, ``Single photon quantum cryptography,'' \prl {\bf 89, } 187901 (2002).

\bibitem{alleaume04} R. All\'{e}aume, F. Treussart, G. Messin, Y. Dumeige, J.-F. Roch, A. Beveratos, R. Brouri-Tualle, J.-P. Poizat, and P. Grangier, ``Experimental open-air quantum key distribution with a single-photon source,'' New\ J.\ Phys.\ {\bf 6, } 92 (2004).

\bibitem{hom87} C. K. Hong, Z. Y Ou, and L. Mandel, ``Measurement of subpicosecond time intervals between two photons by interference,'' \prl {\bf 59, } 2044--2046 (1987).

\bibitem{rohde05} P. P. Rohde, T. C. Ralph, and M. A. Nielsen, ``Optimal photons for quantum-information processing,'' \pra {\bf 72, } 052332 (2005).

\bibitem{turchette95} Q. A. Turchette, C. J. Hood, W. Lange, H. Mabuchi, and H. J. Kimble, ``Measurement of conditional phase shifts for quantum logic,'' \prl {\bf 75, } 4710--4713 (1995).

\bibitem{duan04} L.-M. Duan and H. J. Kimble, ``Scalable photonic quantum computation through cavity-assisted interactions,'' \prl {\bf 92, } 127902 (2004).

\bibitem{duan05} L.-M. Duan, B. Wang, and H. J. Kimble, ``Robust quantum gates on neutral atoms with cavity-assisted photon scattering,'' \pra {\bf 72, } 032333 (2005).

\bibitem{akahane03} Y. Akahane, T. Asano, B.-S. Song, and S. Noda, ``High-Q photonic nanocavity in a two-dimensional photonic crystal,'' \nat {\bf 425, } 944--947 (2003).

\bibitem{song05} B.-S. Song, S. Noda, T. Asano, and Y. Akahane, ``Ultra-high-Q photonic double-heterostructure nanocavity,'' Nature\ Materials\ {\bf 4, } 207--210 (2005).

\bibitem{noda07} S. Noda, M. Fujita, and T. Asano, ``Spontaneous-emission control by photonic crystals and nanocavities,'' Nature\ Photonics\ {\bf 1, } 449--458 (2007).

\bibitem{olivero05} P. Olivero, S. Rubanov, P. Reichart, B. C. Gibson, S. T. Huntington, J. R. Rabeau, A. D. Greentree, J. Salzman, D. Moore, D. N. Jamieson, and S. Prawer, ``Ion-beam-assisted lift-off techniques for three-dimensional micromachining of freestanding single-crystal diamond,'' Advanced\ Materals\ (Weinheim, Ger.)\ {\bf 17, } 2427--2430 (2005).

\bibitem{tomlijenovichanic06} S. Tomljenovic-Hanic, M. J. Steel, C. Martijn de Sterke, and J. Salzman, ``Diamond based photonic crystal microcavities,'' \opex {\bf 14, } 3556--3562 (2006).

\bibitem{baldwin06} J. W. Baldwin, M. Zalalutdinov, T. Feygelson, J. E. Butler, and B. H. Houston, ``Fabrication of short-wavelength photonic crystals in wide-band-gap nanocrystalline diamond films,'' J.\ Vac.\ Sci.\ Technol.\ B\ {\bf 24, } 50--54 (2006).

\bibitem{bayn07} I. Bayn and J. Salzman, ``High-Q photonic crystal nanocavities on diamond for quantum electrodynamics,'' Eur.\ Phys.\ J.\ Appl.\ Phys.\ {\bf 37, } 19--24 (2007).

\bibitem{wang07} C. F. Wang, R. Hanson, D. D. Awschalom, E. L. Hu, T. Feygelson, J. Yang, and J. E. Butler, ``Fabrication and characterization of two-dimensional photonic crystal microcavities in nanocrystalline diamond,'', \apl {\bf 91, } 201112 (2007).

\bibitem{davies76} G. Davies and M. F. Hamer, ``Optical studies of the 1.945eV vibronic band in diamond,'' Proc.\ R.\ Soc.\ Lond.\ A:\ Math.\ and Phys.\ Sci.\ {\bf 348,} 285--298 (1976).

\bibitem{jelezko04} F. Jelezko, T. Gaebel, I. Popa, A. Gruber, and J.Wrachtrup, ``Observation of coherent oscillations in a single electron spin,'' \prl {\bf 92, } 076401 (2004).

\bibitem{tamarat06b} Ph. Tamarat, N. B. Manson, R. L. McMurtie, A. Nitsovtsev, C. Santori, P. Neumann, T. Gaebel, F. Jelezko, P. Hemmer, and J. Wrachtrup, ``The excited state structure of the nitrogen-vacancy center in diamond,'' http://arxiv.org/abs/cond-mat/0610357 (2006).

\bibitem{jaynes63} E. T. Jaynes and F. W. Cummings, ``Comparison of quantum and semiclassical radiation theory with application to the beam maser,'' Proc.\ IEEE\ {\bf 51,} 89--109 (1963).

\bibitem{shore93} B. W. Shore and P. L. Knight, ``The Jaynes-Cummings model,'' \jmo {\bf 40,} 1195--1238 (1993).

\bibitem{law97} C. K. Law and H. J. Kimble, ``Deterministic generation of a bit-stream of single-photon pulses,'' \jmo {\bf 44, } 2067--2074 (1997).

\bibitem{purcell46} E. M. Purcell, ``Spontaneous emission probabilities at radio frequencies,'' Phys.\ Rev.\ {\bf 69,} 681 (1946).

\bibitem{haroche85} S. Haroche and J. M. Raimond, ``Radiative properties of Rydberg states in resonant cavities,'' in \textit{Advances in Atomic and Molecular Physics Vol. XX}, D. Bates and B. Bederson, eds. (Academic, 1985), pp. 350--411.

\bibitem{gerard99} J.-M. G\'{e}rard and B. Gayral, ``Strong Purcell effector for InAs quantum boxes in three-dimensional solid-state microcavities,'' J.\ Lightwave\ Technol.\ {\bf 17, } 2089--2095 (1999).

\bibitem{khanbekyan07} M. Khanbekyan, D.-G. Welsh, C. Di Fidio, and W. Vogel, ``Cavity-assisted spontaneous emission as a single-photon source: Pulse shape and efficiency of one-photon Fock state preparation,'' http://arxiv.org/abs/0709.2998 (2007).

\bibitem{tian92} L. Tian and H. J. Carmichael, ``Quantum trajectory simulations of the two-state behavior of an optical cavity containing one atom,'' \pra {\bf 46, } R6801--R6804 (1992).

\bibitem{carmichael93} H. J. Carmichael, \textit{An Open System Approach to Quantum Optics} (Springer, 1993).

\bibitem{dumeige04} Y. Dumeige, F. Treussart, R. All\'eaume, T. Gacoin, J.-F. Roch, and P. Grangier, ``Photo-induced creaton of nitrogen-related color centers in diamond nanocrystals under femtosecond illumination,'' J.\ Lumin.\ {\bf 109, } 61--67 (2004).

\bibitem{tamarat06a} Ph. Tamarat, T. Gaebel, J. R. Rabeau, M. Khan, A. D. Greentree, H. Wilson, L. C. L. Hollenberg, S. Prawer, P. Hemmer, F. Jelezko, and J. Wrachtrup, ``Stark shift control of single optical centres in diamond,'' \prl {\bf 97, } 083002 (2006).

\bibitem{greentree06c} A. D. Greentree, J. Salzman, S. Prawer, and L. C. L. Hollenberg, ``Quantum gate for Q switching in monolithic photonic-band-gap cavities containing two-level atoms,'' \pra {\bf 73, } 013818 (2006).

\bibitem{fernee07} M. J. Fern\'{e}e, H. Rubinsztein-Dunlop, and G. J. Milburn, ``Improving single-photon sources with Stark tuning,'' \pra {\bf 75, } 043815 (2007).

\bibitem{chang06} D. E. Chang, A. S. S\o rensen, P. R. Hemmer, and M. D. Lukin, ``Quantum optics with surface plasmons,'' \prl {\bf 97, } 053002 (2006).
\end{thebibliography}
\end{document}